\documentclass[a4paper,twocolumn,10pt,accepted=2020-10-04]{quantumarticle}
\pdfoutput=1
\usepackage[utf8]{inputenc}
\usepackage[english]{babel}
\usepackage[T1]{fontenc}
\usepackage{amsmath}
\usepackage{hyperref}

\usepackage{tikz}
\usepackage{lipsum}

\usepackage{amssymb,amsthm,amsfonts}

\begin{document}

\title{Probing nonclassicality with matrices of phase-space distributions}

\author{Martin Bohmann}
	\email{martin.bohmann@oeaw.ac.at}
	\affiliation{Institute for Quantum Optics and Quantum Information - IQOQI Vienna, Austrian Academy of Sciences, Boltzmanngasse 3, 1090 Vienna, Austria}
	\affiliation{QSTAR, INO-CNR, and LENS, Largo Enrico Fermi 2, I-50125 Firenze, Italy}
	\orcid{0000-0003-3857-4555}
\author{Elizabeth Agudelo}
	\affiliation{Institute for Quantum Optics and Quantum Information - IQOQI Vienna, Austrian Academy of Sciences, Boltzmanngasse 3, 1090 Vienna, Austria}
	\orcid{0000-0002-5604-9407}
\author{Jan Sperling}
	\affiliation{Integrated Quantum Optics Group, Applied Physics, Paderborn University, 33098 Paderborn, Germany}
	\orcid{0000-0002-5844-3205}
\maketitle

\begin{abstract}
	We devise a method to certify nonclassical features via correlations of phase-space distributions by unifying the notions of quasipro\-babilities and matrices of correlation functions.
	Our approach complements and extends recent results that were based on Chebyshev's integral inequality [\href{https://doi.org/10.1103/PhysRevLett.124.133601}{Phys. Rev. Lett. \textbf{124}, 133601 (2020)}].
	The method developed here correlates arbitrary phase-space functions at arbitrary points in phase space, inclu\-ding multimode scenarios and higher-order correlations.
	Furthermore, our approach provides necessary and sufficient nonclassicality criteria, applies to phase-space functions beyond $s$-parametrized ones, and is accessible in experiments.
	To demonstrate the power of our technique, the quantum characteristics of discrete- and continuous-variable, single- and multimode, as well as pure and mixed states are certified only employing second-order correlations and Husimi functions, which always resemble a classical probability distribution.
	Moreover, nonlinear generalizations of our approach are studied.
	Therefore, a versatile and broadly applicable framework is devised to uncover quantum properties in terms of matrices of phase-space distributions.
\end{abstract}
\section{Introduction}\label{sec:Introduction}

	Telling classical and quantum features of a physical system apart is a key challenge in quantum physics.
	Besides its fundamental importance, the notion of (quantum-optical) nonclassicality provides the basis for many applications in photonic quantum technology and quantum information \cite{KLM01,RL09a,OFJ09,KMSUZ16,SP19}.
	Nonclassicality is, for example, a resource in quantum networks \cite{YBTNGK18}, quantum metrology \cite{KCTVJ19}, boson sampling \cite{SLR17}, or distributed quantum computing \cite{SLR19}.
	The corresponding free (i.e., classical) operations are passive linear optical transformations and measurement.
	By exceeding such operations, protocols which utilize nonclassical states can be realized.
	Furthermore, nonclassicality is closely related to entanglement.
	Each entangled state is nonclassical, and single-mode nonclassicality can be converted into two- and multi-mode entanglement \cite{KSBK02,VS14,KSP16}.

	Consequently, a plethora of techniques to detect nonclassical properties have been developed, each coming with its own operational meanings for applications.
	For example, quantumness criteria which are based on correlation functions and phase-space representations have been extensively studied in the context of nonclassical light \cite{MBWLN10,SV20}.

	The description of physical systems using the phase-space formalism is one of the cornerstones of modern physics \cite{S01,ZFC05,N10}.
	Beginning with ideas introduced by Wigner and others \cite{W27,W32,G46,M49}, the notion of a phase-space distribution for quantum systems generalizes principles from classical statistical theories (including statistical mechanics, chaos theory, and thermodynamics) to the quantum domain.
	However, the nonnegativity condition of classical probabilities does not translate well to the quantum domain.
	Rather, the notion of quasiprobabilities---i.e., normalised distributions that do not satisfy all axioms of probability distributions and particularly can attain negative values---was established and found to be the eminent feature that separates classical concepts from genuine quantum effects.
	(See Refs. \cite{SW18,SV20} of thorough introductions to quasiproabilities.)

	In particular, research in quantum optics significantly benefited from the concept of phase-space quasiprobability distributions, including prominent examples, such as the Wigner function \cite{W32}, the Glauber-Sudarshan $P$ function \cite{G63,S63}, and the Husimi $Q$ function \cite{H40}.
	In fact, the very definition of nonclassicality---the impossibility of describing the behaviour of quantum light with classical statistical optics---is directly connected to negativities in such quasiprobabilities, more specifically, the Glauber-Sudarshan $P$ function \cite{TG65,M86}.
	Because of the general success of quasiprobabilities, other phase-space distributions for light have been conceived \cite{C66,CG69,AW70}, each coming with its own advantages and drawbacks.
	For example, squeezed states are represented by nonnegative (i.e., classical) Wigner functions although they form the basis for continuous-variable quantum information science and technology \cite{BL05,WPGCRSL12,ARL14}, also having a paramount role for quantum metrology \cite{GDDSSV13,T19}.

	Another way of revealing nonclassical effects is by using correlation constraints which, when violated, witness nonclassicality.
	Typically, such conditions are formulated in terms of inequalities involving expectation values of different observables.
	Examples in optics are photon anti-bunching \cite{CW76,KM76,KDM76} and sub-Poissonian photon-number distributions \cite{M79,ZM90}, using intensity correlations, as well as various squeezing criteria, being based on field-operator correlations \cite{Y76,W83,LK87,A93}.
	They can follow, for instance, from applying Cauchy-Schwartz inequalities \cite{A88} and uncertainty relations \cite{H87}, as well as from other violations of classical bounds \cite{K96,RL09,BQVC19}.
	Remarkably, many of these different criteria can be jointly described via so-called matrix of moments approaches \cite{AT92,SV05a,SV06,MPHH09,SV06b}.
	However, each of the mentioned kinds of nonclassicality, such as squeezed and sub-Poissonian light, requires a different (sub-)matrix of moments, a hurdle we aim at overcoming.

	Over the last two decades, there had been many attempts to unify matrix-of-moment-based criteria with quasiprobabilities.
	For example, the Fourier transform of the $P$ function can be used, together with Bochner's theorem, to correlate such transformed phase-space distributions through determinants of a matrix \cite{V00,RV02}, being readily available in experimental applications \cite{LS02,ZPB07,KVHDSS09,MKNPE11}, and further extending to the Laplace transformation \cite{SVA16}.
	Furthermore, a joint description of field-operator moments and transformed phase-space functions has been investigated as well \cite{RSAMKHV15}.
	Rather than considering matrices of phase-space quasiproabilities, concepts like a matrix-valued distributions enable us to analyzed nonclassical hybrid systems \cite{WMV97,ASCBZV17}.
	Very recently, a first successful strategy that truly unifies correlation functions and phase-space functions has been conceived \cite{BA19}.
	However, these first demonstrations of combining phase-space distributions and matrices of moments are still restricted to rather specific scenarios.

	In this contribution, we formulate a general framework for uncovering quantum features through correlations in phase-space matrices which unifies these two fundamental approaches to characterizing quantum systems.
	By combining matrix of moments and quasiprobabilities, this method enables us to probe nonclassical characteristics in different points in phase space, even using different phase-space distributions at the same time.
	We specifically study implications from the resulting second- and higher-order phase-space distribution matrices for single- and multimode quantum light.
	Furthermore, a direct measurement scheme is proposed and non-Gaussian phase-space distributions are analyzed.
	To benchmark our method, we consider a manifold of examples, representing vastly different types of quantum features.
	In particular, we show that our matrix-based approach can certify nonclassicality even if the underlying phase-space distribution is nonnegative.
	In summary, our approach renders it possible test for nonclassicality by providing easily accessible nonclassicality conditions.
    While previously derived phase-space-correlation conditions \cite{BA19} were restricted to single-mode scenarios, the present approach straightforwardly extends to multimode cases.
    In addition, our phase-space matrix technique includes nonclassicality-certification approaches based on phase-space distributions and matrices of moments as special cases, resulting in an overarching structure that combines both previously separated techniques.

	The paper is structured as follows.
	Some initial remarks are provided in Sec. \ref{sec:Prelim}.
	Our method is rigorously derived and thoroughly discussed in Sec. \ref{sec:PSM}.
	Section \ref{sec:GenImp} concerns several generalizations and potential implementations of our toolbox.
	Various examples are analyzed in Sec. \ref{sec:Ex}.
	Finally, we conclude in Sec. \ref{sec:Conclusion}.

\section{Preliminaries}\label{sec:Prelim}

	In their seminal papers \cite{G63,S63}, Glauber and Sudarshan showed that all quantum states of light can be represented diagonally in a coherent-state basis through the Glauber-Sudarshan $P$ distribution.
	Specifically, a single-mode quantum state can be expanded as
	\begin{align}
		\label{eq:GSrepresentation}
		\hat \rho=\int d^{2}\alpha\, P(\alpha)|\alpha\rangle\langle\alpha|,
	\end{align}
	where $|\alpha\rangle$ denotes a coherent state with a complex amplitude $\alpha$.
	Then, classical states are identified as statistical (i.e., incoherent) mixtures of pure coherent states, which resemble the behavior of a classical harmonic oscillator most closely \cite{S26,H85}.
	For this diagonal representation to exist for nonclassical states as well, the Glauber-Sudarshan distribution has to exceed the class of classical probability distributions \cite{TG65,M86}, particularly violating the nonnegativity constraint, $P\ngeq 0$.
	This classification into states which have a classical correspondence and those which are genuinely quantum is the common basis for certifying nonclassical light.

    As laid out in the introduction, nonclassicality is a vital resource for utilizing quantum phenomena, ranging from fundamental to applied \cite{YBTNGK18,KCTVJ19,SLR17,SLR19}.
    In this context, it is worth adding that, contrasting other notions of quantumness, nonclassicality is based on a classical wave theory.
    That is, it is essential to discern nonclassical coherence phenomena from those which are accessible with classical statistical optics, as formalized through Eq. \eqref{eq:GSrepresentation} with $P\geq0$.
    See, e.g., Ref. \cite{RSG19} for a recent experiment that separates classical and quantum interference effects in such a manner.
    For instance, free operations, i.e., those maps which preserve classical states, do include beam splitter transformations, resulting in the generation of entanglement from single-mode nonclassical states via such a free operation \cite{KSBK02,VS14,KSP16} that is vital for many quantum protocols.

\subsection{Phase-space distributions}

	Since the Glauber-Sudarshan distribution can be a highly singular distribution (see, e.g., Ref. \cite{S16}), generalized phase-space functions have been devised.
	Within the wide range of quantum-optical phase-space representations, the family of $s$-parametrized distributions \cite{CG69,AW70} is of particular interest.
	Such distributions can be expressed as
	\begin{align}
		\label{eq:RelExpPSD}
		P(\alpha;\sigma)
		=\frac{\sigma}{\pi}
		\left\langle {:}\exp\left(
		    -\sigma\hat n(\alpha)
		\right){:}\right\rangle,
	\end{align}
	where colons indicate normal ordering \cite{VW06} and $\hat n(\alpha)=(\hat a-\alpha)^\dagger (\hat a-\alpha)$ is the displaced photon-number operator, written in terms of bosonic annihilation and creation operators, $\hat a$ and $\hat a^\dag$, respectively.
	It is worth recalling that the normal ordering acts on the expression surrounded by the colons in such a way that creation operators are arranged to the left of annihilation operators whilst ignoring commutation relations.
	Note that, for convenience, we parametrize distributions via the width parameter $\sigma$, rather than using $s$.
	Both are related via
	\begin{equation}
		\label{eq:sTOsigma}
		\sigma=\frac{2}{1-s}.
	\end{equation}
	From this relation, we can identify the Husimi function, $Q(\alpha)=P(\alpha;1)$, for $s=-1$ and $\sigma=1$; the Wigner function, $W(\alpha)=P(\alpha;2)$, for $s=0$ and $\sigma=2$; and the Glauber-Sudarshan function,  $P(\alpha)=P(\alpha;\infty)$, for $s=1$ and $\sigma=\infty$.

	Whenever a phase-space distribution contains a negative contribution, i.e., $P(\alpha;\sigma)<0$ for at least one pair $(\alpha;\sigma)$, the underlying quantum state is nonclassical \cite{TG65,M86}.
	In such a case, the distribution $P(\alpha;\sigma)$ refers to as a quasiprobability distribution which is incompatible with classical probability theory.
	Nonetheless, for any $\sigma\geq0$ and any state, this function represents a real-valued distribution which is normalized, $P(\alpha;\sigma)=P(\alpha;\sigma)^\ast$ and $\int d^2\alpha\, P(\alpha;\sigma)=1$.
	In addition, it is worth mentioning that the normalization of the state is guaranteed through the limit
	\begin{equation}
		\lim_{\sigma\to 0}\frac{\pi}{\sigma} P(\alpha;\sigma)=\langle{:}\exp(0){:}\rangle=\langle\hat 1\rangle=1.
	\end{equation}

\subsection{Matrix of moments approach}

	Besides phase-space distributions, a second family of nonclassicality criteria is based on correlation functions; see, e.g., Refs. \cite{SRV05,SV05} for introductions.
	For this purpose, we can consider an operator function $\hat f=f(\hat a, \hat a^\dagger)$.
	Then, 
	\begin{align}
		\label{eq:fdagf}
		\langle{:}\hat f^\dagger\hat f{:}\rangle
		=\int d^2\alpha\, P(\alpha)|f(\alpha,\alpha^*)|^2
		\stackrel{\text{cl.}}{\geq}0
	\end{align}
	holds true for all $P\geq 0$.
	Now, one can expand $\hat f$ in terms of a given set of operators, e.g., $\hat f=\sum_{i} c_i \hat O_i$, resulting in $\langle{:}\hat f^\dag\hat f{:}\rangle=\sum_{i,j} c_i^\ast c_j \langle {:}\hat O_i^\dag\hat O_j{:}\rangle$.
	Furthermore, this expression is nonnegative [cf. Eq. \eqref{eq:fdagf}] iff the matrix $(\langle {:}\hat O_i^\dag\hat O_j{:}\rangle)_{i,j}$ is positive semidefinite.
	This constraint can, for example, be probed using Sylvester's criterion \cite{HJ90} which states that a Hermitian matrix is positive-definite if and only if all its leading principal minors have a positive determinant.
	It is worth mentioning that Eq. \eqref{eq:fdagf} defines the notion of a nonclassicality witness, where $\langle{:}\hat f^\dagger\hat f{:}\rangle<0$ certifies nonclassicality.

	The above observations form the basis for many experimentally accessible nonclassicality criteria, such as using basis operators which are powers of quadrature operators \cite{AT92}, photon-number operators \cite{A93}, and general creation and annihilation operators \cite{SRV05,SV05}.
	See Refs. \cite{MBWLN10} for an overview of moment-based inequalities.
	In the following, we are going to combine the phase-space distribution technique with the method of matrices of moments to arrive at the sought-after unifying approach of both techniques.

\section{Matrix of phase-space distributions}\label{sec:PSM}

	Both phase-space distributions and matrices of moments exhibit a rather dissimilar structure when it comes to formulating constraints for classical light.
	Consequently, a full unification of both approaches is missing to date, excluding the few attempts mentioned in Sec. \ref{sec:Introduction}.
	In this section, we bridge this gap and derive a matrix of phase-space distributions which leads to previously unknown nonclassicality criteria, also overcoming the limitations of earlier methods.

\subsection{Derivation}

	For the purpose of deriving our criteria, we consider an operator function $\hat f=\sum_i c_i \exp[-\sigma_i\hat n_i(\alpha_i)]$.
	Then, the normally ordered expectation value of $\hat f^\dag\hat f$ can be expanded as
	\begin{align}
		\label{eq:QuadraticForm}
	\begin{aligned}
		&\langle{:}\hat f^\dagger \hat f{:}\rangle
		=\sum_{i,j} c_i^\ast c_j\langle{:}e^{-\sigma_i\hat n(\alpha_i)}e^{-\sigma_j\hat n(\alpha_j)}{:}\rangle
		\\
		=&\sum_{i,j} c_j^\ast c_i
		\exp\left[-\frac{\sigma_i\sigma_j}{\sigma_i+\sigma_j}|\alpha_i-\alpha_j|^2\right]
		\\ &\times
		\left\langle{:}\exp\left[
		-(\sigma_i+\sigma_j)\,\hat n \left(\frac{\sigma_i\alpha_i+\sigma_j\alpha_j}{\sigma_i+\sigma_j}\right)
		\right]{:}\right\rangle.
	\end{aligned}
	\end{align}
	Based on the above relation, we may define two matrices,
	one for classical amplitudes,
	\begin{align}
		M^\mathrm{(c)}=\left(
		\exp\left[-\frac{\sigma_i\sigma_j}{\sigma_i+\sigma_j}|\alpha_i-\alpha_j|^2\right]
		\right)_{i,j},
	\end{align}
	and one for the quantum-optical expectation values,
	\begin{align}
	\begin{aligned}
		M^\mathrm{(q)}=&\left(
			\left\langle{:}\exp\left[
				-(\sigma_i+\sigma_j)\,\hat n \left(\frac{\sigma_i\alpha_i+\sigma_j\alpha_j}{\sigma_i+\sigma_j}\right)
			\right]{:}\right\rangle
			\right)_{i,j}
		\\ =&
		\left(\frac{\pi}{\sigma_i+\sigma_j}\,P\left(\frac{\sigma_i\alpha_i+\sigma_j\alpha_j}{\sigma_i+\sigma_j};\sigma_i+\sigma_j\right)\right)_{i,j},
	\end{aligned}
	\end{align}
	which can be expressed in terms of phase-space distributions using Eq. \eqref{eq:RelExpPSD}.
	Specifically, $M^\mathrm{(q)}$ corresponds to a matrix of phase-space distributions.
	Moreover, the fact that the normally ordered expectation value of $\hat f^\dag\hat f$ is nonnegative for classical light [Eq. \eqref{eq:fdagf}] is then identical to the entry-wise product (i.e., the Hadamard product $\circ$) of both matrices being positive semidefinite,
	\begin{align}
		M\stackrel{\text{cl.}}{\geq}0,
		\text{ with }
		M=M^\mathrm{(c)}\circ M^\mathrm{(q)}
	\end{align}
	defining our phase-space matrix $M$.

	For classical light, all principal minors of $M$ have to be nonnegative according to Sylvester's criterion.
	Conversely, the violation of this constraint certifies a nonclassical state,
	\begin{align}
		\label{eq:NclPSM}
		\det(M)<0,
	\end{align}
	where $M$ is defined through arbitrary small or large sets of parameters $\sigma_i$ and $\sigma_j$ and coherent amplitudes $\alpha_i$ and $\alpha_j$.
	Therefore, inequality \eqref{eq:NclPSM} enables us to formulate various nonclassicality conditions which correlate distinct phase-space distributions as it typically only done for matrix-of-moments-based techniques when using different kinds of observables.
	We finally remark that the expression in Eq. \eqref{eq:NclPSM} resembles a nonlinear nonclassicality witnessing approach.

	As a first example, we may explore the first-order criterion, i.e., a $1\times 1$ matrix of quasiprobabilities.
	Selecting arbitrary $\sigma$-parameters and coherent amplitudes, i.e., ($\alpha_1;\sigma_1)=(\alpha;\sigma)$, we find the following restriction for classical states [cf. Eq. \eqref{eq:NclPSM}]:
	\begin{align}\label{eq:1x1}
		\frac{\pi}{2\sigma}P(\alpha;2\sigma)\stackrel{\text{cl.}}{\geq}0.
	\end{align}
	This inequality reflects the fact that finding negativities in a parametrized phase-space distribution $P(\alpha;2\sigma)$ is sufficient to certify nonclassicality.
	Also recall that we retrieve the Glauber-Sudarshan distribution in the limit $\sigma\to\infty$.
	Since the nonnegativity of this distribution defines the very notion of a nonclassical state \cite{TG65,M86}, we can conclude from this examples that our approach is necessary and sufficient for certifying nonclassicality.

	However, the Glauber-Sudarshan distribution has the disadvantage of being a highly singular for many relevant nonclassical states of light and, thus, hard to reconstruct from experimental data.
	Consequently, it is of practical importance (see Secs. \ref{sec:GenImp} and \ref{sec:Ex}) to consider higher-order criteria beyond this trivial one.

\subsection{Second-order criteria}

	We begin our consideration with an interesting second-order case.
	We chose $(\alpha_1;\sigma_1)=(0;0)$ and $(\alpha_2;\sigma_2)=(\alpha;\sigma)$.
	This yields the $2\times 2$ phase-space matrix
	\begin{align}
		\label{eq:2x2specific}
		M=\begin{pmatrix}
			1 & \langle{:}\exp(-\sigma \hat n(\alpha)){:}\rangle
			\\ \langle{:}\exp(-\sigma \hat n(\alpha)){:}\rangle & \langle{:}\exp(-2\sigma \hat n(\alpha)){:}\rangle
		\end{pmatrix}.
	\end{align}
	Up to a positive scaling, the determinant of this matrix results in the following nonclassicality criterion:
	\begin{equation}
		\label{eq:WQ}
		P(\alpha;2\sigma)
		-\frac{2\pi}{\sigma}\left(P(\alpha;\sigma)\right)^2
		< 0.
	\end{equation}
	In particular, we can set $\sigma=1$ to relate this condition to the Wigner and Husimi functions, leading to $W(\alpha)-2\pi Q(\alpha)^2<0$.
	This special case of our general approach has been recently derived using a very different approach, using Chebyshev's integral inequality \cite{BA19}.
	There it was shown that, by applying the inequality \eqref{eq:WQ} for $\sigma=1$, it is possible to certify nonclassicality even if the Wigner function of the state under study is nonnegative.
	In this context, remember that the Husimi function, $Q(\alpha)=\langle\alpha|\hat\rho|\alpha\rangle/\pi$, is always nonnegative, regardless of the state $\hat\rho$.

	Beyond this scenario, we now study a more general $2\times 2$ phase-space matrix $M$.
	For an efficient description, it is convenient to redefine transformed parameters as
	\begin{subequations}
	\begin{eqnarray}
		\label{eq:TrafoAmp}
		\Delta\alpha = \alpha_2-\alpha_1
		&\text{ and }&
		A = \frac{\sigma_1\alpha_1+\sigma_2\alpha_2}{\sigma_1+\sigma_2},
		\\ \label{eq:TrafoS}
		\tilde\sigma = \frac{\sigma_1\sigma_2}{\sigma_1+\sigma_2}
		&\text{ and }&
		\Sigma = \sigma_1+\sigma_2.
	\end{eqnarray}
	\end{subequations}
	Note that these parameters relate to the two-body problem.
	That is, the quantities in Eq. \eqref{eq:TrafoAmp} define the relative position and barycenter in phase-space, respectively; and the two quantities in Eq. \eqref{eq:TrafoS} resemble the reduced and total mass in a mechanical system, respectively.

	In this alternative parametrization, the two matrices, giving the total phase-space matrix $M=M^\mathrm{(c)}\circ M^\mathrm{(q)}$, read
	\begin{subequations}
	\begin{align}
		M^\mathrm{(c)}=&\begin{pmatrix}
		1 & e^{-\tilde\sigma|\Delta\alpha|^2} 
		\\
		e^{-\tilde\sigma|\Delta\alpha|^2}  & 1
		\end{pmatrix}, \quad \text{and}
		\\
		M^\mathrm{(q)}=&\begin{pmatrix}
		\langle{:} e^{-2\sigma_1\hat n(\alpha_1)} {:}\rangle
		& \langle{:} e^{-\Sigma\hat n\left(A\right)} {:}\rangle
		\\
		\langle{:} e^{-\Sigma\hat n\left(A\right)} {:}\rangle
		& \langle{:} e^{-2\sigma_2\hat n(\alpha_2)} {:}\rangle
		\end{pmatrix}.
	\end{align}
	\end{subequations}
	Hence, the determinant of the Hadamard product of both matrices then gives
	\begin{align}
		\label{eq:2x2}
		\det(M) =
		\langle{:} e^{-2\sigma_1\hat n(\alpha_1)} {:}\rangle
		\langle{:} e^{-2\sigma_2\hat n(\alpha_2)} {:}\rangle\nonumber\\
		{-} e^{-2\tilde\sigma|\Delta\alpha|^2}
		\langle{:} e^{-\Sigma\hat n\left(A\right)} {:}\rangle^2.
	\end{align}
	If this determinant is negative for the state of light under study, its nonclassicality is proven.
	In terms of phase-space distributions, this condition can be also recast into the form
	\begin{align}
		\label{eq:2x2condition}
		P(\alpha_1;2\sigma_1)P(\alpha_2;2\sigma_2)
		-\frac{4\tilde\sigma}{\Sigma}\left[e^{-\tilde\sigma|\Delta\alpha|^2}P(A;\Sigma)\right]^2<0.
	\end{align}
	Interestingly, this nonclassicality criterion correlates different points in phase space for different distributions, $P(\alpha_1;2\sigma_1)$ and $P(\alpha_2;2\sigma_2)$, with a phase-space distribution with the total width $\Sigma$ at the barycenter $A$ of coherent amplitudes, $P(A;\Sigma)$.

\subsection{Higher-order cases}\label{sec:HigherOrderCases}

	The next natural extension concerns the analysis of higher-order correlations.
	Clearly, one can obtain an increasingly large set of nonclassicality tests with an increasing dimensionality of $M$, determined by the number of pairs $(\alpha_i;\sigma_i)$.
	In order to exemplify this potential, let us focus on one specific $3\times 3$ scenario and more general scenarios for specific choices of parameters.

\begin{widetext}
	Let us discuss the $3\times 3$ case firstly, for which we are going to consider $\sigma_3=0$.
	From this, one obtains the following phase-space matrix:
	\begin{equation}
		M=\begin{pmatrix}
			\langle{:}\exp(-2\sigma_1\hat n(\alpha_1)){:}\rangle & \langle{:}\exp(-\sigma_1\hat n(\alpha_1)-\sigma_2\hat n(\alpha_2)){:}\rangle & \langle{:}\exp(-\sigma_1\hat n(\alpha_1)){:}\rangle
			\\
			\langle{:}\exp(-\sigma_1\hat n(\alpha_1)-\sigma_2\hat n(\alpha_2)){:}\rangle & \langle{:}\exp(-2\sigma_2\hat n(\alpha_2)){:}\rangle & \langle{:}\exp(-\sigma_2\hat n(\alpha_2)){:}\rangle
			\\
			\langle{:}\exp(-\sigma_1\hat n(\alpha_1)){:}\rangle & \langle{:}\exp(-\sigma_2\hat n(\alpha_2)){:}\rangle & 1
		\end{pmatrix}.
	\end{equation}
	Again, directly expressing this matrix in terms of phase-space functions, as done previously, we get a third-order nonclassicality criterion from its determinant  \cite{comment:3x3}.
	It reads
	\begin{equation}
	\begin{aligned}
		 \frac{\det(M)}{\pi^2} =&
		\left(
			\frac{P(\alpha_1;2\sigma_1)}{2\sigma_1}
			-\pi\left(\frac{P(\alpha_1;\sigma_1)}{\sigma_1}\right)^2
		\right)
		\left(
			\frac{P(\alpha_2;2\sigma_2)}{2\sigma_2}
			-\pi\left(\frac{P(\alpha_2;\sigma_2)}{\sigma_2}\right)^2
		\right)
		\\ & -
		\left(
			\exp(-\tilde\sigma|\Delta\alpha|^2)\frac{P(A;\Sigma)}{\Sigma}
			-\pi\frac{P(\alpha_1;\sigma_1)}{\sigma_1}\frac{P(\alpha_2;\sigma_2)}{\sigma_2}
		\right)^2<0,
	\end{aligned}
	\end{equation}
	using the parameters defined in Eqs. \eqref{eq:TrafoAmp} and \eqref{eq:TrafoS}.
	In fact, this condition combines the earlier derived criteria of the forms \eqref{eq:WQ} and \eqref{eq:2x2} in a manner similar to cross-correlations nonclassicality conditions known from matrices of moments \cite{SVA16}.
\end{widetext}

	Another higher-order matrix scenario corresponds to having identical coherent amplitudes, i.e., $\alpha_i=\alpha$ for all $i$.
	In this case, we find that the two Hadamard-product components of the matrix $M$ simplify to
	\begin{align}
	\begin{aligned}
		M^\mathrm{(c)}&=(1)_{i,j}\quad
		\text{and}\quad\\
		M^\mathrm{(q)}&=\left(
		\frac{\pi}{\sigma_i+\sigma_j} P\left(\alpha;\sigma_i+\sigma_j\right)
		\right)_{i,j},
	\end{aligned}
	\end{align}
	thus resulting in $M=M^\mathrm{(q)}$.
	Therefore, we can formulate nonclassicality criteria which correlate an arbitrary number of different phase-space distributions, defined via $\sigma_i$, at the same point in phase-space, $\alpha$.

	Analogously, one can consider a scenario in which all $\sigma$ parameters are identical, $\sigma_i=\sigma$.
	Then, we get
	\begin{align}
	\begin{aligned}
		M^\mathrm{(c)}=&(e^{-\sigma|\alpha_i-\alpha_j|^2/2})_{i,j}
		\qquad\text{and}
		\\
		M^\mathrm{(q)}=&\left(
		\frac{\pi}{2\sigma} P\left(\frac{\alpha_i+\alpha_j}{2};2\sigma\right)
		\right)_{i,j}.
	\end{aligned}
	\end{align}
	Consequently, we obtain nonclassicality criteria which correlate an arbitrary number of different points in phase space, $\alpha_i$, for a single phase-space distribution, parametrized by $\sigma$.

\subsection{Comparison with Chebyshev's integral inequality approach}

	As mentioned previously, a related method based on Chebyshev's integral inequality has been introduced recently \cite{BA19}.
	It also provides inequality conditions for different phase-space distributions.
	The nonclassicality conditions based on Chebyshev's integral inequality take the form
	\begin{align}
		\label{eq:Chebyshev}
		P(\alpha;\Sigma)-\frac{\Sigma}{\pi}\prod_{i=1}^{D} \left[\frac{\pi}{\sigma_i} P(\alpha;\sigma_i)\right]<0,
	\end{align}
	where $\Sigma=\sum_{i=1}^{D} \sigma_i$.
	To compare both approaches, let us discuss their similarities and differences.

	In its simplest form, involving only $\sigma_1$ and $\sigma_2$, the condition in  Eq. \eqref{eq:Chebyshev} resembles the tests based on the $2\times2$ matrix in Eq. \eqref{eq:2x2specific}.
	In particular, for the case $\sigma_1=\sigma_2=\sigma$ both methods yield the exact same conditions.
	For $\sigma_1\neq\sigma_2$ such an agreement of both methods cannot be found because of the inherent symmetry of the phase-space matrix approach, $M=M^\dag$, which stems from its construction via a quadratic form; cf. Eq. \eqref{eq:QuadraticForm}.
	Also, for more general, higher-order conditions, i.e. $D>2$, such similarities cannot be found either.
	Conditions of the form in Eq. \eqref{eq:Chebyshev} consist of only two summands.
	The first term is a single phase-space function with the width parameter $\Sigma$ which is associated to the highest $\sigma$ parameter involved in the inequality.
	The second term is a product of $D$ phase-space distributions, each individual distribution has a width parameter $\sigma_i$, together bound by the condition $\Sigma=\sum_{i=1}^{D} \sigma_i$.
	By comparison, our phase-space matrix approach yields, in general, a richer and more complex set of higher-order nonclassicality tests, such as demonstrated in Sec. \ref{sec:HigherOrderCases}.

	Let us point out further differences between the two approaches.
	Firstly, we observe that the inequalities based on Chebyshev's integral inequality only apply to one single point in phase space.
	In contrast, the phase-space matrix method devised here includes conditions that combine different points in phase space; cf. Eq. \eqref{eq:QuadraticForm}.
	Secondly, Chebyshev's integral inequality approach cannot be extended to multimode settings.
	Such a limitation does not exist for the matrix approach either, as we show in the following Sec. \ref{subsec:multimode}.
	We conclude that both the technique in Ref. \cite{BA19} and our phase-space matrix approach for obtaining phase-space inequalities yield similar second-order conditions but, in general, give rise to rather different nonclassicality criteria.
	In particular, the phase-space matrix framework offers a broader range of variables---be it coherent amplitudes or widths---that lead to a richer set of nonclassicality conditions.

\subsection{Extended relations to nonclassicality criteria}

    To finalize our first discussions we now focus on the relation to matrices of moments.
    Previously, we have shown that, already in the first order, our criteria are necessary and sufficient to verify the nonclassicality, and we discussed our method in relation to Chebychev's integral inequality.
    Furthermore, indirect techniques using transformed phase-space functions, such as the characteristic function \cite{RSAMKHV15} and the two-sided Laplace transform \cite{SVA16}, have been previously related to moments.
    Thus, the question arises what the relation of our direct technique to such matrices of moments is.

    For showing that our framework includes the matrix of moments technique, we may remind ourselves that derivatives can be understood as a linear combination, specifically as a limit of a differential quotient, $\partial_z^mg(z)=\lim_{\epsilon\to0}\epsilon^{-m}\sum_{k=0}^m\binom{m}{k}(-1)^{m-k}g(z+k\epsilon)$.
    This enables us to write \cite{comment:derivatives}
    \begin{equation}
        \hat a^{\dag m}\hat a^{n}
        =\sigma^{-(m+n)}\left.\partial_{\alpha}^m\partial^n_{\alpha^\ast} e^{\sigma|\alpha|^2}{:}e^{-\sigma\hat n(\alpha)}{:}\right|_{\alpha=0 \text{ and }\sigma=0},
    \end{equation}
    expressing arbitrary moments $\hat a^{\dag m}\hat a^{n}$ via linear combinations of the normally ordered operators that represent $\sigma$-parametrized phase-space distributions.
    Thus, in the corresponding limits, we can identify the operator $\hat f$ in Eq. \eqref{eq:fdagf} with $\hat f=\sum_{m,n}c_{m,n}\sigma^{-(m+n)}\partial_{\alpha}^m\partial^n_{\alpha^\ast} e^{\sigma|\alpha|^2}{:}e^{-\sigma\hat n(\alpha)}{:}|_{\alpha=0,\sigma=0}=\sum_{m,n}c_{m,n}\hat a^{\dag m}\hat a^{n}$.
    For such a choice $\hat f$, $\langle{:}\hat f^\dag\hat f{:}\rangle<0$ is in fact identical to the most general form of the matrix of moments criterion for nonclassicality \cite{SRV05,SV05}.

    In conclusion, we find that our necessary and sufficient methodology not only includes nonclassicality criteria based on phase-space functions themselves [cf. Eq. \eqref{eq:1x1}], but it also includes the technique of matrices of moments as a special case.
    In a hierarchical picture, this means that our family of nonclassicality criteria, including arbitrary orders of $\sigma$-parametrized phase-space functions, encompasses both negativities of phase-space functions and matrices of moments.
    Because of the above relation, the order of moments that is required to certify nonclassicality also sets an upper bound to the size of the matrix of phase-space distributions so that it certifies nonclassicality.
    Therefore, our approach unifies and subsumes both earlier types of nonclassicality conditions.

\section{Generalizations and implementation}\label{sec:GenImp}

	In this section, we generalize our approach to arbitrary multimode nonclassical light and propose a measurement scheme to experimentally access the matrix of phase-space distributions.
	In addition, we show that our approach applies to phase-space distributions which are no longer limited to $\sigma$ parametrizations and relate these findings to the response of nonlinear detection devices.

\subsection{Multimode case}\label{subsec:multimode}

	After our in-depth analysis of single-mode phase-space matrices, the multimode case follows almost straightforwardly.
	For the purpose of such a generalization, we consider $N$ optical modes, represented via the annihilation operators $\hat a_m$ for $m=1,\ldots,N$ and extending to the displaced photon-number operators $\hat n_m(\alpha^{(m)})=(\hat a_m-\alpha^{(m)})^\dag(\hat a_m-\alpha^{(m)})$.
	Now, $\sigma$-parametrized multimode phase-space functions can be expressed as
	\begin{equation}
	\begin{aligned}
		&\left\langle{:}
			e^{-\sigma^{(1)}\hat n_1(\alpha^{(1)})}
			\cdots
			e^{-\sigma^{(N)}\hat n_N(\alpha^{(N)})}
		{:}\right\rangle
		\\
		=& \frac{\pi^N}{\sigma^{(1)}\cdots \sigma^{(N)}}
		P(\alpha^{(1)},\ldots,\alpha^{(N)} ; \sigma^{(1)},\ldots,\sigma^{(N)}),
	\end{aligned}
	\end{equation}
	where we allow for different $s$ parameters for each mode, with $s^{(m)}=1-2/\sigma^{(m)}$ [Eq. \eqref{eq:sTOsigma}].
	As in the single-mode case, we can now formulate a matrix $M$ of multimode phase-space functions,
	\begin{align*}
		M{=}\left(\left\langle{:}
			e^{-\sum\limits_{m=0}^N\sigma_i^{(m)}\hat n_m(\alpha_i^{(m)})}
			e^{-\sum\limits_{m=0}^N\sigma_j^{(m)}\hat n_m(\alpha_j^{(m)})}
		{:}\right\rangle\right)_{i,j}.
	\end{align*}
	Consequently, this matrix of phase-space functions also has to be positive semidefinite if the underlying state of multimode light is classical.
	That is,
    \begin{equation}
        M\stackrel{\text{cl.}}{\geq}0
    \end{equation}
    holds true for classical light and for any dimension (or order) of the multimode matrix $M$ and any sigma and alpha values.
    Conversely, $\det(M)<0$ is a nonlinear witness of multimode nonclassicality.
    Similarly to the single-mode case, an increasingly large matrix $M$ with increasingly dense sets of parameters for the various alpha and sigma values then enables one to probe the nonclassicality of arbitrary multimode states.

	Since we have already exemplified various scenarios for single-mode phase-space correlations, in the following, we restrict ourselves to a particular multimode case.
	Specifically, we focus on two optical modes and a $3\times 3$ phase-space matrix $M$ is,
	\begin{equation}
		\begin{pmatrix}
			1 & \frac{\pi P(\alpha^{(1)};\sigma)}{\sigma} & \frac{\pi P(\alpha^{(2)};\sigma)}{\sigma}
			\\
			\frac{\pi P(\alpha^{(1)};\sigma)}{\sigma} & \frac{\pi P(\alpha^{(1)};2\sigma)}{2\sigma} & \frac{\pi^2P(\alpha^{(1)},\alpha^{(2)};\sigma,\sigma)}{\sigma^2}
			\\
			\frac{\pi P(\alpha^{(2)};\sigma)}{\sigma} & \frac{\pi^2P(\alpha^{(1)},\alpha^{(2)};\sigma,\sigma)}{\sigma^2} & \frac{\pi P(\alpha^{(1)};2\sigma)}{2\sigma}
		\end{pmatrix},\nonumber
	\end{equation}
	where quasiprobabilities as a function of single-mode parameters indicate marginal phase-space distributions.
	Adopting a notation of pairs of coherent amplitudes and widths, $M$ is thus defined via the following two-mode parameters: $(\alpha^{(1)}_1,\alpha^{(2)}_1;\sigma^{(1)}_1,\sigma^{(2)}_1)=(0,0;0,0)$, $(\alpha^{(1)}_2,\alpha^{(2)}_2;\sigma^{(1)}_2,\sigma^{(2)}_2)=(\alpha^{(1)},0;\sigma,0)$, and $(\alpha^{(1)}_3,\alpha^{(2)}_3;\sigma^{(1)}_3,\sigma^{(2)}_3)=(0,\alpha^{(2)};0,\sigma)$.
	In particular, we can express the nonclassicality constraint from the determinant of $M$ \cite{comment:3x3} for $\sigma=1$ via joint and marginal Wigner and Husimi functions,
	\begin{align}
	\begin{aligned}
		\frac{\det M}{\pi^4}
		=& \left[
			\tfrac{W(\alpha^{(1)})}{2\pi}-Q(\alpha^{(1)})^2
		\right]\left[
			\tfrac{W(\alpha^{(2)})}{2\pi}-Q(\alpha^{(2)})^2
		\right]
		\\ &
		-\left[
			Q(\alpha^{(1)},\alpha^{(2)})-Q(\alpha^{(1)})Q(\alpha^{(2)})
		\right]^2 \stackrel{\text{cl.}}{\geq}0.
	\end{aligned}
	\end{align}
	Violating this inequality verifies the nonclassicality of the two-mode state under study.

\subsection{Direct measurement scheme}\label{subsec:Detect}

\begin{figure}
	\includegraphics[width=.95\columnwidth]{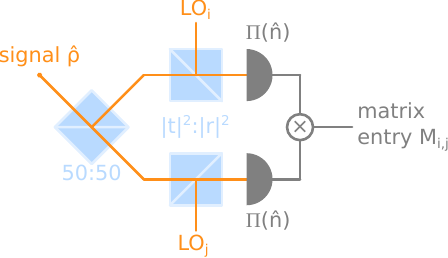}
	\caption{
		Outline of phase-space matrix correlation measurement.
		The signal, i.e., the state $\hat\rho$ of the light field under study, is split into two identical outputs at a 50:50 beam splitter.
		Each of the resulting beams is combined with a local oscillator (LO) on a $|t|^2:|r|^2$ beam splitter and measured with a photon-number-based detector, represented through $\Pi(\hat n)$.
		The resulting correlations yield the entries of our phase-space matrix $M$.
	}\label{fig:setup}
\end{figure}

	The reconstruction of phase-space distributions can be a challenging task \cite{LR09}.
	For this reason, we are going to devise a directly accessible setup to infer the phase-space matrix.
	See Fig. \ref{fig:setup} for an outline which is based on the approaches in Refs. \cite{WV96,SV05,BRWK99}.
	For convenience, we restrict ourselves to a single optical mode; the extension to multiple modes follows straightforwardly.
	That is, each of the multiple modes can be detected individually by a correlation-measurement setup as depicted in Fig. \ref{fig:setup}.
	Furthermore, it is noteworthy that our phase-space matrix approach is not limited to this specific measurement scheme proposed here and generally applies to any detection scenario which allows for a reconstruction of quasiprobability distributions.

	For the setup in Fig. \ref{fig:setup}, we begin our considerations with a coherent state $|\beta\rangle$, representing our signal $\hat\rho=|\beta\rangle\langle \beta|$.
	Firstly, we split this signal equally into $2$ modes, resulting in a two-mode coherent state $|\beta/\sqrt2,\beta/\sqrt2\rangle$.
	In addition, local oscillator states are prepared, $|\beta_i\rangle$ and $|\beta_j\rangle$ for each mode.
	Each of the two signals is then mixed with its local oscillator on a $|t|^2{:}|r|^2$ beam splitter, where $|t|^2+|r|^2=1$.
	One output of each beam splitter is discarded, namely the lower and upper one for the top and bottom path in Fig. \ref{fig:setup}, respectively.
	This results in the input-output relation
	\begin{align}
		|\beta\rangle\mapsto \left|t\frac{\beta}{\sqrt2}+r\beta_i,t\frac{\beta}{\sqrt2}+r\beta_j\right\rangle,
	\end{align}
	which is then detected as follows.

	Each of the resulting modes is measured with a detector or detection scheme based on photon absorption, thus being described by a positive operator-valued measure (POVM) which is diagonal in the photon-number representation \cite{KK64}.
	Consequently, one or a combination of detector outcomes (e.g., in a generating-function-type combination \cite{SPBTEWLNLGVASW20})  corresponds to a POVM element of the form $\Pi(\hat n)={:}e^{-\Gamma(\hat n)}{:}$.
	Using $|m\rangle\langle m|={:}e^{-\hat n}\hat n^m/m!{:}$ for an $m$-photon projector, this means that we identify $\sum_{m=0}^\infty\pi_m|m\rangle\langle m|={:}e^{-\hat n}\sum_{m=0}\pi_m \hat n^m/m!{:}=\Pi(\hat n)={:}e^{-\Gamma(\hat n)}{:}$, where the eigenvalues $\pi_m$ corresponds to the Taylor expansion coefficients of the function $z\mapsto\exp[z-\Gamma(z)]$.
	Accordingly, the function $\Gamma(\hat n)$ models the detector response \cite{KK64,VW06}.
	Finally, the correlation measurement of this response for our coherent signal states takes the form
	\begin{equation}
		\label{eq:DetectCohState}
	\begin{aligned}
		M_{i,j}&=
		\exp\left(
			-\Gamma\left(\frac{|t|^2}{2}\left|
				\beta - \frac{r\sqrt 2\beta_i}{t}
			\right|\right)
		\right)
		\\ &\times
		\exp\left(
			-\Gamma\left(\frac{|t|^2}{2}\left|
				\beta + \frac{r\sqrt 2\beta_j}{t}
			\right|\right)
		\right).
	\end{aligned}
	\end{equation}

	Now it is convenient to define $\tilde\Gamma(\hat n)=\Gamma(|t|^2\hat n/2)$ and
	\begin{equation}
		\alpha_i=-\frac{r\sqrt 2\beta_{i}}{t},
	\end{equation}
	for all LO choices $i$ and, similarly, for $j$.
	Furthermore, we generalize this treatment to arbitrary states, $\hat\rho=\int d^2\beta\, P(\beta)|\beta\rangle\langle\beta|$, using the Glauber-Sudarshan representation [Eq. \eqref{eq:GSrepresentation}].
	Therefore, the correlations measured as described above [Eq. \eqref{eq:DetectCohState}] obey
	\begin{equation}
		M_{i,j}=\left\langle{:}
			e^{-\tilde\Gamma(\hat n(\alpha_i))}
			e^{-\tilde\Gamma(\hat n(\alpha_j))}
		{:}\right\rangle,
	\end{equation}
	which corresponds to a directly measured phase-space matrix element, e.g., for a linear detector response $\tilde\Gamma(\hat n)=\sigma \hat n$.
	The other way around, we can choose $\hat f=\sum_{i} c_i\exp(-\tilde\Gamma(\hat n(\alpha_i)))$ for the general classicality constraint in Eq. \eqref{eq:fdagf}, even for nonlinear detector responses.
	Then, the matrix of phase-space distribution approach applies, regardless of a linear or nonlinear detection model.
	(See also Refs. \cite{ASW93,SPBTEWLNLGVASW20} in this context.)

	As an example, we consider a case with two single on-off click detectors (represented by $\Pi(\hat n)$ in Fig. \ref{fig:setup}) with a non-unit quantum efficiency $\eta_{\det}$ and a non-vanishing dark-count rate $\delta$ \cite{comment:darkcount}, which represents realistic detectors in experiments.
	In addition, we can introduce neutral density (ND) filters to attenuate the light that impinges on each detector.
	The POVM element for the no-click event in combination with the ND filters then reads $\hat\Pi(\hat n)={:}\exp(-(\eta\hat n+\delta)){:}$, where $0\leq \eta\leq\eta_{\det}$ is a controllable efficiency.
	The measured correlation for this scenario takes the form
	\begin{equation}
		M_{i,j}=\exp(-2\delta)\langle{:}\exp\left(
			-\eta_i\hat n(\alpha_i)-\eta_j\hat n(\alpha_j)
		\right){:}\rangle.
	\end{equation}
	Therein, the adjustable efficiency $\eta_i$ plays role of $\sigma_i$.
	Also, the positive factor that includes the dark counts is irrelevant because it does not change the sign of the determinant of $M$, i.e., the verified nonclassicality.

	In summary, the measurement layout in Fig. \ref{fig:setup} enables us to directly measure the entries of our phase-space matrix $M$.
	As an experimental setup, this scheme also underlines the strong connection between correlations and their measurements and phase-space quasiprobabilities and their reconstruction.
	We may emphasize that all experimental techniques and components that are used in the proposed setup are readily available; see, e.g., the related quantum state reconstruction experiments reported in Refs. \cite{BTBSSV18,SPBTEWLNLGVASW20}.

\subsection{Generalized phase-space functions}\label{subsec:Regularized}

	The $\sigma$-parametrized phase-space distributions we considered so far are related to each other via convolutions with Gaussian distributions \cite{C66,CG69,AW70}.
	However, there are additional means to represent a state without relying on Gaussian convolutions only.
	Such generalized phase-space function can be obtained from the Glauber-Sudarshan $P$ function via
	\begin{equation}
		P_\Omega(\alpha)
		=\int d^2\tilde\alpha\, P(\tilde \alpha)\,\Omega(\alpha;\tilde\alpha,\tilde\alpha^\ast)
		=\langle{:}\Omega(\alpha;\hat a,\hat a^\dag){:}\rangle
	\end{equation}
	for a kernel $\Omega\geq0$ \cite{AW70,KV10}.
	The construction of this so-called filter or regularizing function $\Omega$ can be done so that the resulting distribution $P_\Omega$ is regular (i.e., without the singular behavior known from the $P$ function) and is positive semidefinite for all classical states \cite{KV10}.
	For instance, a non-Gaussian filter $\Omega$ has been used to experimentally characterize squeezed states via regular distributions which exhibit negativities in phase space \cite{KVHS11};
	this cannot be done with $s$-parametrized quasiprobability distributions, which are either nonnegative or highly singular for squeezed states.

	As done for the previously considered distributions, we can define an operator $\hat f=\sum_i c_i \Omega_i(\alpha;\hat a,\hat a^\dag)$, which leads to a phase-space matrix with the entries
	\begin{equation}
		M_{i,j}=\langle{:}\Omega_i(\alpha;\hat a,\hat a^\dag)\Omega_j(\alpha;\hat a,\hat a^\dag){:}\rangle=P_{\Omega_i\Omega_j}(\alpha).
	\end{equation}
	This expression utilizes product of filters $\Omega(\alpha;\tilde\alpha,\tilde\alpha^\ast)=\Omega_i(\alpha;\tilde\alpha,\tilde\alpha^\ast)\Omega_j(\alpha;\tilde\alpha,\tilde\alpha^\ast)$ to be convoluted with the $P$ function.
	From this definition of a regularized phase-space matrix, we can proceed as we did earlier to formulate nonclassicality criteria in terms of phase-space functions.

	Moreover, the non-Gaussian filter functions can be even related to nonlinear detectors.
	For this purpose, we assume that $\Omega(\alpha;\tilde\alpha,\tilde\alpha^\ast)=\Omega(|\alpha-\tilde\alpha|^2)$ (likewise, ${:}\Omega(\alpha,\hat a,\hat a^\dag){:}={:}\Omega(\hat n(\alpha)){:}$ in the normally ordered operator representation).
	In this form, the function is invariant under rotations.
	As we did for the general POVM element $\Pi(\hat n)$, we can now identify
	\begin{equation}
		\Gamma(\hat n)=-\ln\Omega(\hat n).
	\end{equation}
	This enables us to associate non-Gaussian filters and nonlinear detectors and, by extension, generalized phase-space matrices for certifying nonclassical states of light.
	An example for this treatment is studied in Sec. \ref{subsec:ExNL}.

\section{Examples and benchmarking}\label{sec:Ex}

	In the following, we apply our method of phase-space matrices to various examples and benchmark its performance.
	For the latter benchmark, we could consider different phase-space functions.
	Using the $P$ function would be impractical as it is often a highly singular distribution.
	The Wigner function is regular and can exhibit negativities.
	But error estimations from measured data can turn out to be rather difficult because it requires diverging pattern functions \cite{R96,LMKRR96} (see Ref. \cite{SVKMH15} for an in-depth analysis).
	Beyond those practical hurdles, we focus on the $Q$ function here because, already in theory, it is always nonnegative.
	Thus, it is hard to verify nonclassical features based on this particular phase-space distribution.
	Additionally, the $Q$ function is easily accessible in experiments and can be directly measured via the widely-used double-homodyne (aka, eight-port homodyne) detection scheme \cite{VW06}.

	Nonetheless, we are going to demonstrate that, with our method, it is already sufficient for many examples to consider second-order correlations of $Q$ functions.
	For this purpose, we use the condition in Eq. \eqref{eq:2x2condition}, which follows from the $2\times2$ matrix condition with $\sigma_1=\sigma_2=1/2$.
	This special case of that condition then reads as 
	\begin{align}
		\label{eq:QQ}
		\det(M)=Q(\alpha_1)Q(\alpha_{2})
		-e^{-|\alpha_2-\alpha_{1}|^2/2}Q\left(\tfrac{\alpha_1+\alpha_2}{2}\right)^2<0.
	\end{align}
	Meaning that, when the correlations from $Q$ functions at different points in phase space fall below the classical limit zero, nonclassical light is certified with the nonnegative family of $Q$ distributions.

	Moreover, since $Q$ functions are nonnegative, the second term in Eq. \eqref{eq:QQ} is subtractive in nature.
	Thus, it is sufficient to find a point $\alpha_1$ in phase space for which $Q(\alpha_1)=0$ holds true---together with an $\alpha_2$ with $Q(\alpha_2/2)>0$, which has to exist because of normalization---in order to certify nonclassicality through Eq. \eqref{eq:QQ}.
	Setting $\alpha_1=\alpha$, this leads to the simple nonclassicality condition $Q(\alpha)=0$, which applies to arbitrary quantum states.
	In Ref. \cite{LB95},  this specific condition has been independently verified as a nonclassical signature of non-Gaussian states.
	Here, we see that this nonclassical signature is indeed a corollary of our general approach.
	Furthermore, we remark that this condition only holds if the $Q$ function is exactly zero.
	In experimental scenarios, in which errors have to be accounted for, it is infeasible to get this exact value.
	Therefore, the condition Eq. \eqref{eq:QQ} is more practical as it allows us to certify nonclassicality through a finite negative value.
	Furthermore, this condition is applicable even if $Q(\alpha)=0$ does not hold true.

\subsection{Discrete-variable states}

	We start our analysis of nonclassicality by considering discrete-variable states for a single mode.
	In the case of quantized harmonic oscillators, such as electromagnetic fields, a family of discrete-variable states that are of particular importance are number states $|n\rangle$.
	They represent an $n$-fold excitation of the underlying quantum field and show the particle nature of said fields, thus being nonclassical when compared to classical electromagnetic wave phenomena.
	However, photon-number states require Glauber-Sudarshan $P$ distributions that are highly singular because they involve up to $2n$th-order derivatives of delta distributions \cite{VW06}.
	On the other hand, the $Q$ function of photon-number states,
	\begin{align}
		\label{eq:PhotonQ}
		Q_{|n\rangle}(\alpha)=\frac{|\alpha|^{2n}}{\pi n!} e^{-|\alpha|^2},
	\end{align}
	is an accessible and smooth, but nonnegative function.
	Thus, by itself, it cannot behave as a quasiprobability which includes negative contributions that uncover nonclassicality.

\begin{figure}
	\includegraphics[width=0.95\columnwidth]{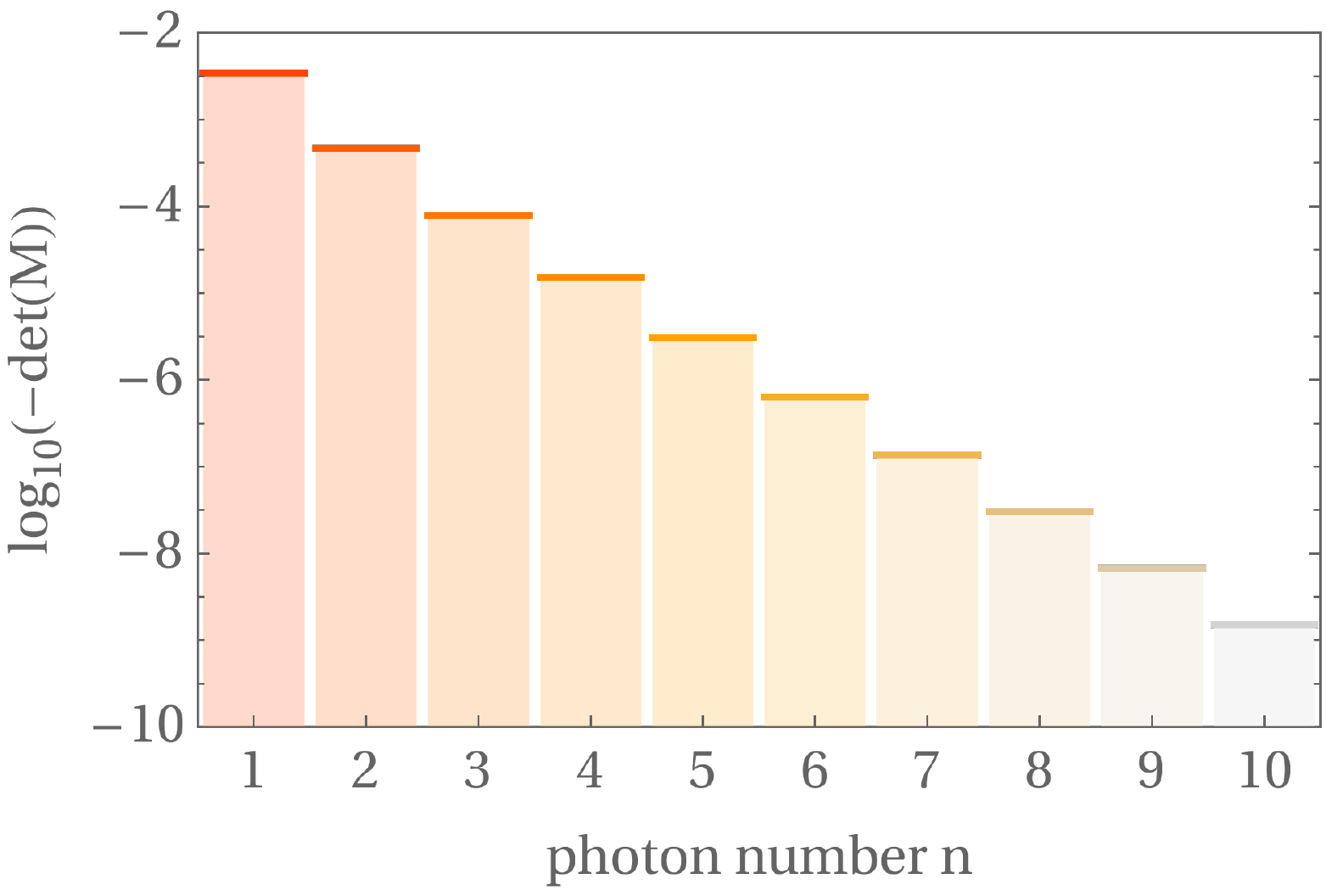}
	\caption{
		Nonclassicality of number states $|n\rangle$ via Eq. \eqref{eq:QQ} on a logarithmic scale.
		We choose $\alpha_1=0$ and determined an optimal $\alpha_2=\sqrt{2n}$ as points in phase space to correlate $Q$ functions.
		The largest certification of nonclassicality is found for a single photon, $n=1$, and it decreases thereafter.
	}\label{fig:QQFock}
\end{figure}

	Except for vacuum, the $Q_{|n\rangle}$ function is zero for $\alpha=0$ and positive for all other arguments $\alpha$ [Eq. \eqref{eq:PhotonQ}].
	Consequently, we can apply Eq. \eqref{eq:QQ} with $\alpha_1=0$ and $\alpha_2\neq0$, yielding $\det(M)<0$.
	Furthermore, a straightforward optimization shows that $|\alpha_2|=\sqrt{2n}$ results in the minimal value $\det(M)=-e^{-2n}(n/2)^{2n}/(\pi n!)^2$.
	Note that this family of discrete-variable number states is rotationally invariant, rendering the phase of $\alpha_2$ irrelevant.
	In Fig. \ref{fig:QQFock}, we visualize the results of our analysis.
	For all number states, we observe a successful verification of nonclassicality in terms of inequality Eq. \eqref{eq:QQ}.
	The single-photon state shows the largest violation for this specific nonclassicality test, and the negativity of $\det(M)$ decreases with the number of photons.
	A possible explanation for this behavior is that this condition is most sensitive towards the particle nature of the quantum states, being most prominent in the single excitation of the quantized radiation field.
	Again, let us emphasize that we verified nonclassciality via a matrix $M$ of classical (i.e., nonnegative) phase-space functions.

\subsection{Continuous-variable states}

	After studying essential examples of discrete-variable quantum states, we now divert our attention to typical examples of continuous-variable states.
	For this reason, we consider squeezed vacuum states which are defined as $|\xi\rangle=(\cosh{r})^{-1/2}\sum_{n=0}^\infty (-e^{i\varphi}\tanh[r]/2)^n\sqrt{(2n)!}|2n\rangle/n!$, for a squeezing parameter $r=|\xi|$ and a phase $\varphi=\arg(\xi)$.
	Without a loss of generality, we set $\varphi=0$.
	Squeezed states are widely used in quantum optical experiments and provide the basis of continuous-variable quantum information processing \cite{BL05}.
	Their parametrized phase-space distributions are known to be either highly singular or nonnegative Gaussian functions (see, e.g., Refs. \cite{WPGCRSL12,S16}).
	For example, the $Q$ function of the states under study can be written as 
	\begin{align}
		\label{eq:Qsqueezed}
		Q_{|\xi\rangle}(\alpha)=\frac{
			\exp\left[-|\alpha|^2-\tanh(r)\mathrm{Re}(\alpha^2)\right]
		}{\pi \cosh (r)}.
	\end{align}
	In the context of earlier discussions, note that this $Q$ function is not zero for $\alpha=0$, or anywhere else.

\begin{figure}
	\includegraphics[width=0.95\columnwidth]{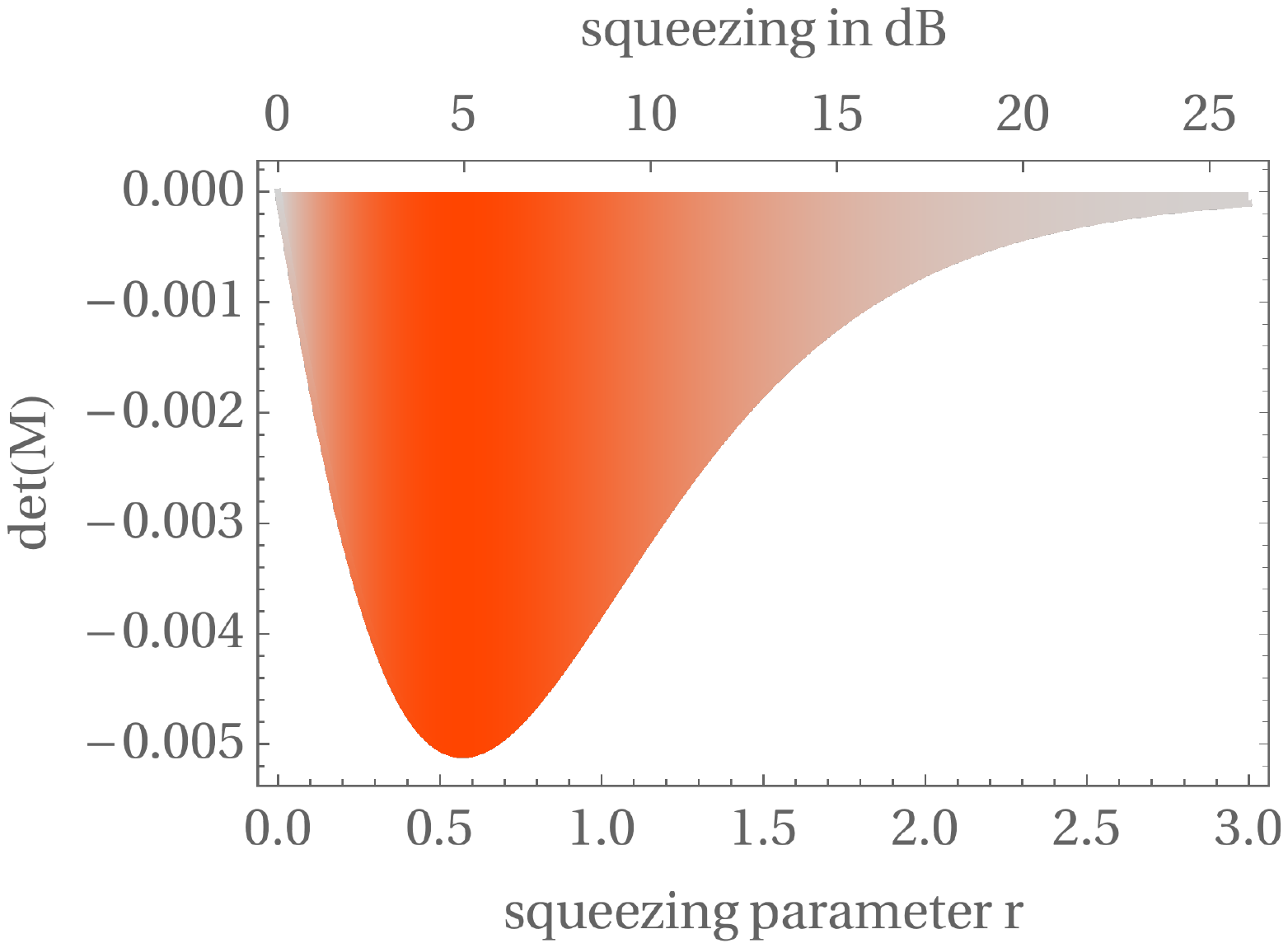}
	\caption{
		The maximally negative value for inequality \eqref{eq:QQ} as a function of the squeezing parameter $r$ is depicted, for the choice $\alpha_1=0$.
		Because of $\det(M)<0$, nonclassicality is certified via $Q$ functions for all $r\neq 0$ [with $Q(\alpha)\gneqq0$ for all $\alpha$].
		A maximal violation of the classical constraint $\det(M)\geq0$ for the considered $2\times2$ matrix $M$ is found for $4.95\,\mathrm{dB}$ [$=10\,\log_{10}(e^{-2 r})$ for $r\approx0.57$] squeezing.
	}\label{fig:SqueezedQ}
\end{figure}

	In Fig. \ref{fig:SqueezedQ}, the left-hand side of inequality Eq. \eqref{eq:QQ} is shown for the $Q_{|\xi\rangle}$ function of a squeezing parameter $r$.
	The points in phase space are determined by choosing $\alpha_1=0$ and minimizing $\det(M)$, being solved for $\alpha_2=[(2/\lambda)\ln[(1+\lambda)/(1+ \lambda/2)]]^{1/2}$, where $\lambda=\tanh(r)$.
	We observe negative values as a direct signature of the nonclassicality of squeezed states.
	Remarkably, this is achieved using the same criterion that applies to photon-number states, typically vastly different correlation functions are required (using either photon numbers \cite{M79} or quadratures \cite{Y76}).
	While inequality Eq. \eqref{eq:QQ} is violated for any squeezing parameter $r>0$, we see that there exists an optimal region of squeezing values around $r=0.6$ (likewise, $5\,\mathrm{dB}$ of squeezing) for which the considered criterion is optimal.
	In particular, this shows that this condition works optimally in a range of moderate squeezing values and, thus, is compatible with typical experiments.
	We also want to recall that the $Q_{|\xi\rangle}$ are a Gaussian distributions which do not have any zeros in the phase space.
	Thus, criteria based on the zeros of the Husimi $Q$ function \cite{LB95} cannot detect nonclassicality in this scenario.
	In contrast, our inequality condition can even certify this Gaussian nonclassicality, hence providing a more sensitive approach to detecting quantum light.

\subsection{Mixed two-mode states}

\begin{figure}
	{\large (a)}\phantom{MMMMMMMMMMMMMMMMMMMMMMMMMMMMM}\\
	\includegraphics[width=0.95\columnwidth]{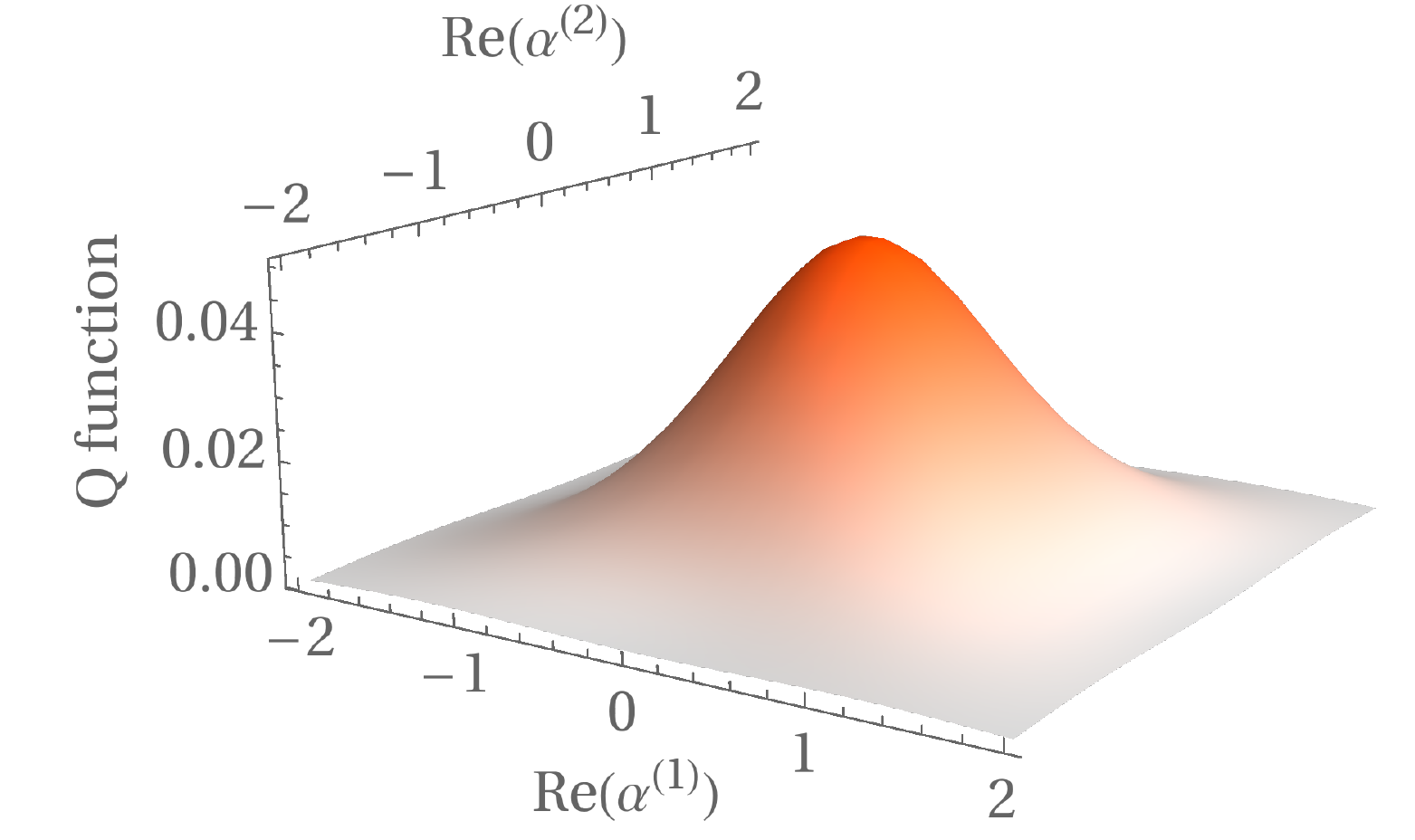}
	{\large (b)}\phantom{MMMMMMMMMMMMMMMMMMMMMMMMMMMMM}\\
	\includegraphics[width=0.95\columnwidth]{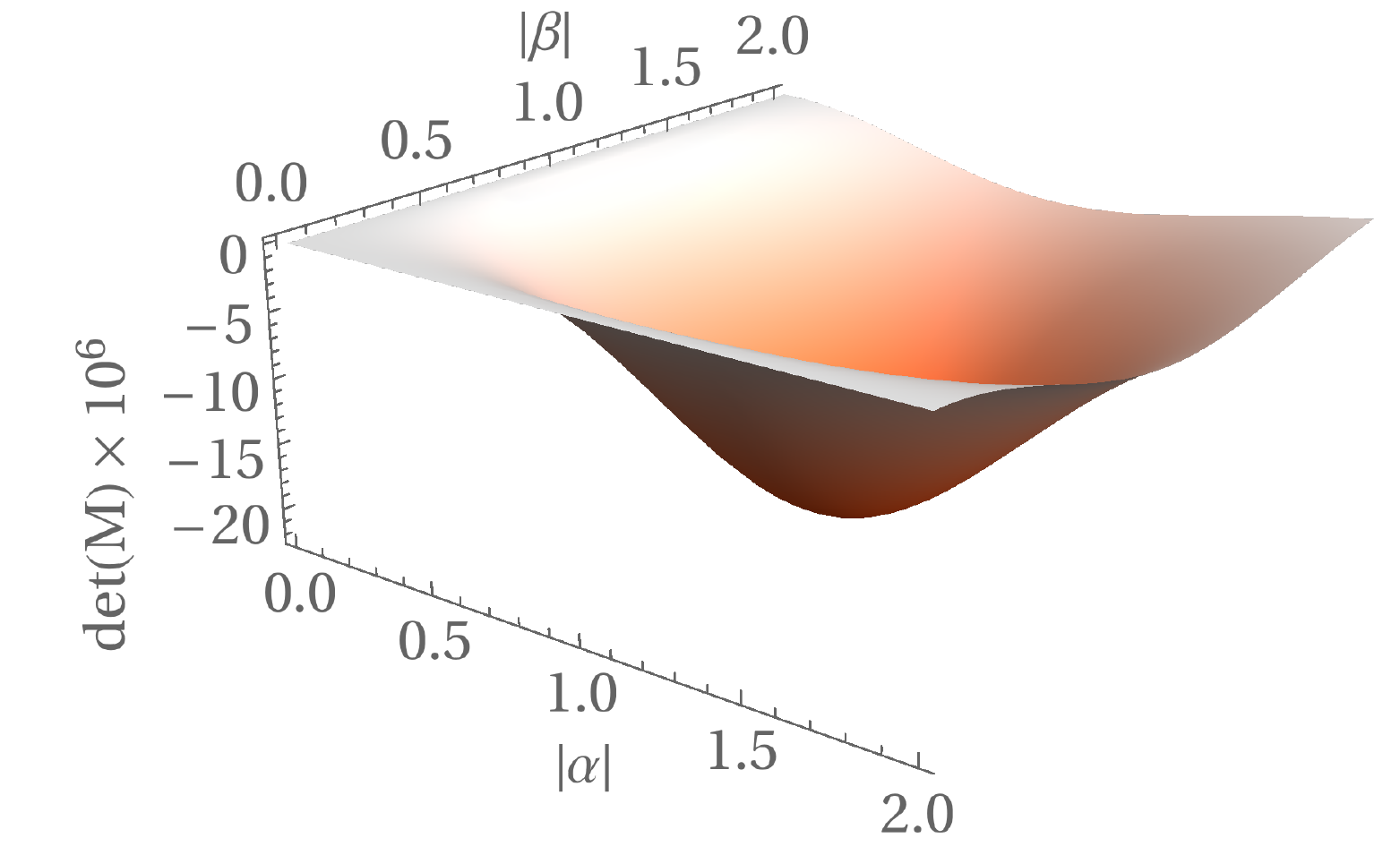}
	\caption{
		In plot (a), the two-mode $Q$ function in Eq. \eqref{eq:Qbipart} for the mixed and weakly correlated state $\hat\rho$ is depicted for $|\lambda|^2=1/2$ and phase-space points with $\mathrm{Im}(\alpha^{(1)})=\mathrm{Im}(\alpha^{(2)})=0$.
		Part (b) visualizes the application of the nonclassicality inequality \eqref{eq:QQmulti} to this state for $N=2$ modes and for the parameter pairs $(\alpha^{(1)}_1,\alpha^{(2)}_1)=(\alpha,0)$ and $(\alpha^{(1)}_2,\alpha^{(2)}_2)=(0,\beta)$.
		Nonclassicality is verified because of $\det(M)<0$, and maximized for $|\alpha|$ and $|\beta|$ around one.
	}\label{fig:ElizaState}
\end{figure}

	To further challenge our approach, we now consider a bipartite mixed state.
	We begin with a two-mode squeezed vacuum state, $|\lambda\rangle=\sqrt{1-|\lambda|^2}\sum_{n=0}^\infty\lambda^n|n,n\rangle$.
	This state undergoes a full phase diffusion, leading to the mixed state
	\begin{equation}
	    \label{Eq:State}
	\begin{aligned}
		\hat\rho&=\frac{1}{2\pi}\int\limits_{0}^{2\pi}d\varphi\, |\lambda e^{i\varphi}\rangle\langle \lambda e^{i\varphi}|
		\\
		&=\sum_{n=0}^\infty (1{-}|\lambda|^2)|\lambda|^{2n} |n,n\rangle\langle n,n|.
	\end{aligned}
	\end{equation}
	This state presents a particular challenge for nonclassicality verification because it shows only weak nonclassicality and quantum correlations.
	Namely, this state is not entangled, has zero quantum discord, and has classical marginal single-mode states (i.e., the partial traces $\mathrm{tr}_1(\hat\rho)=\mathrm{tr}_2(\hat\rho)$ yield thermal states) \cite{ASV13}.
	However, it shows nonclassical photon-photon correlations \cite{FP12,ASV13,SBVHBAS15}.
	The state's two-mode $Q$ function can be computed using Gaussian functions and the phase averaging in Eq. \eqref{Eq:State}, which gives
	\begin{align}
		\label{eq:Qbipart}
	\begin{aligned}
		Q_{\hat\rho}(\alpha^{(1)},\alpha^{(2)})
		= \frac{1{-}|\lambda|^2}{\pi^2}
		e^{-|\alpha^{(1)}|^2-|\alpha^{(2)}|^2}\\
		\times\,
		I_0(2|\lambda||\alpha^{(1)}||\alpha^{(2)}|),
	\end{aligned}
	\end{align}
	where $I_0$ denotes the zeroth modified Bessel function of the first kind.
	See also Fig. \ref{fig:ElizaState}(a) in this context.

	To apply our approach, whilst using $Q$ functions only, we can directly generalize our criterion in Eq. \eqref{eq:QQ} to the multimode case (see also Sec. \ref{subsec:multimode}).
	For $N$ modes, this results in the nonclassicality criterion
	\begin{align}
		\label{eq:QQmulti}
	\begin{aligned}
		\det(M)
		=& Q(\alpha^{(1)}_{1},\ldots,\alpha^{(N)}_{1}) Q(\alpha^{(1)}_{2},\ldots,\alpha^{(N)}_{2})
		\\
		& -e^{-\sum_{m=1}^N |\alpha^{(m)}_2-\alpha^{(m)}_1|^2/2}
		\\
		&\times Q\left(\frac{\alpha^{(1)}_1{+}\alpha^{(1)}_2}{2},\ldots,\frac{\alpha^{(N)}_1{+}\alpha^{(N)}_2}{2}\right)^2<0.
	\end{aligned}
	\end{align}
	
	In Fig. \ref{fig:ElizaState}(b), we apply the case $N=2$ of this inequality to identify the nonclassicality of $\hat\rho$ for $|\lambda|^2=1/2$.
	Again, the same approach as used in both single-mode scenarios enables us yet again to uncover the nonclassical behavior of this bipartite state for all nonzero choices of parameters $|\alpha|$ and $|\beta|$.
	Note in this context that the phase of these parameters does not contribute because of the fully phase-randomized structure of the mixed state in Eq. \eqref{Eq:State}.
	
	\begin{figure}
	\centering
	\includegraphics[width=0.78\columnwidth]{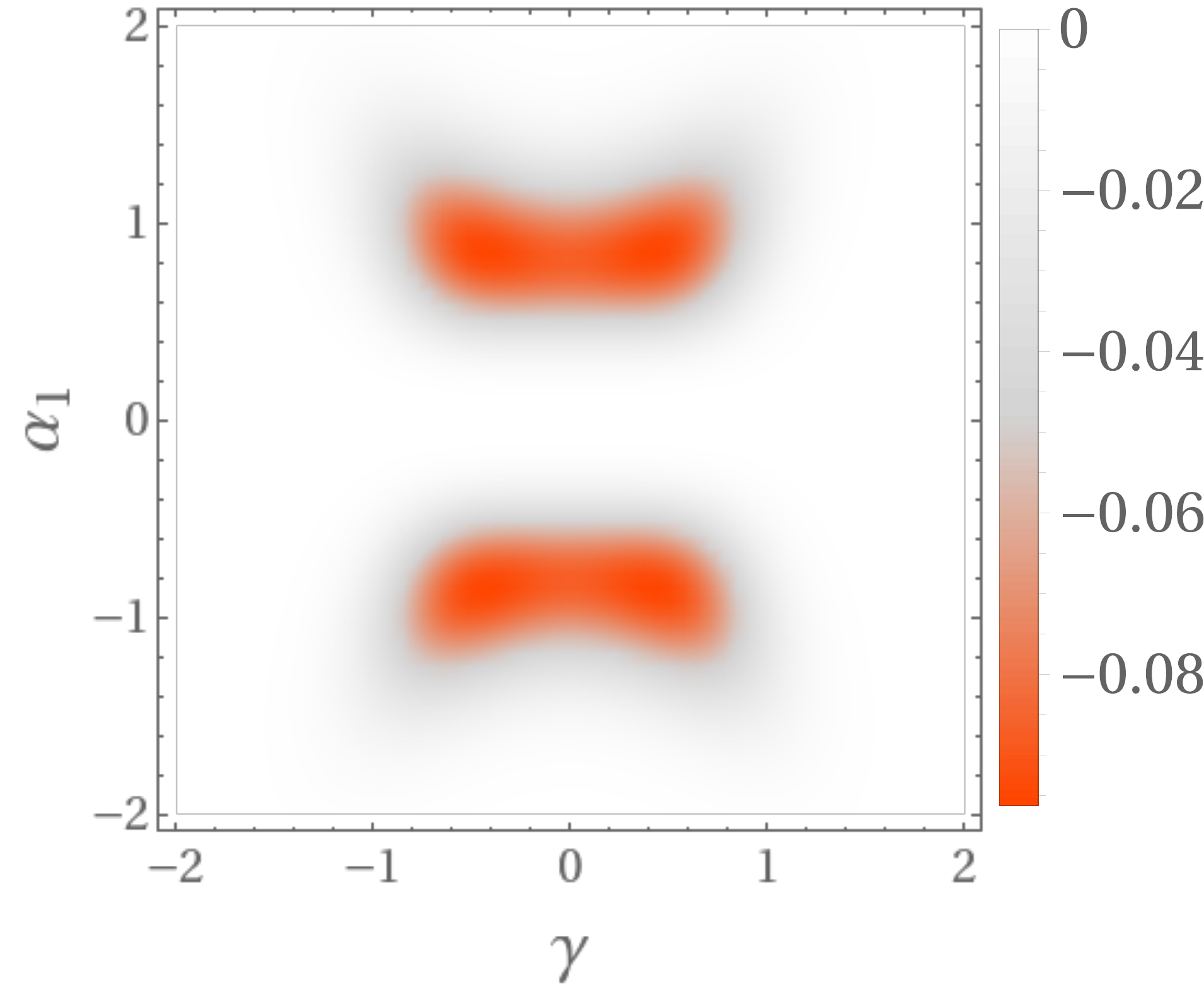}
	\caption{
		Determinant ($\times 10^4$) of the multimode $2\times2$ phase-space matrix $M$ of $Q$ functions [$\det(M)$ in Eq. \eqref{eq:QQmulti}] for the skew-symmetric, tripartite state $|\Psi^{(-)}_{\gamma,3}\rangle$, with $\alpha^{(m)}_1=\alpha_1$ and $\alpha^{(m)}_2=0$ for $m=1,2,3$.
		Nonclassicality is verified for all coherent amplitudes $\gamma$, which, without loss of generality, can be chosen as a nonnegative number.
	}\label{fig:OddCoh}
    \end{figure}

\subsection{Multimode superposition states}

	To further exceed the previous, bipartite state, we consider an $N$-mode state in this part.
	Specifically, we focus on a multimode superposition of coherent states \cite{SMM74},
	\begin{align}
		\label{eq:MultimodeState}
		|\Psi^{(\pm)}_{\gamma,N}\rangle=\frac{|\gamma\rangle^{\otimes N}\pm|-\gamma\rangle^{\otimes N}}{\sqrt{2\left(1\pm e^{-2N|\gamma|^2}\right)}},
	\end{align}
	which consists of two $N$-fold tensor products of polar opposite coherent states, $|\pm\gamma\rangle$.
	Specifically, the skew-symmetric state $|\Psi^{(-)}_{\gamma,N}\rangle$ is of interest because it yields a GHZ state for $|\gamma|\to\infty$ and W state for $|\gamma|\to0$, combining in an asymptotic manner two inequivalent forms of multipartite entanglement \cite{DVC00,SV20}.
	
	\begin{widetext}
	The $Q$ functions for the states in Eq. \eqref{eq:MultimodeState} can be straightforwardly computed; they read
	\begin{align}
	\begin{aligned}
		Q_{|\Psi^{(\pm)}_{\gamma,N}\rangle}(\alpha^{(1)},\ldots,\alpha^{(N)})
		=
		\frac{e^{-N|\gamma|^2} e^{-|\alpha^{(1)}|^2}\cdots e^{-|\alpha^{(N)}|^2}}{2\pi^N\left[1\pm e^{-2N|\gamma|^2}\right]}
		\!\!\left(
			\!\!\cosh\!\!\left[2\,\mathrm{Re}\!\!\left(\!\!\gamma^\ast\sum_{m=1}^N \alpha^{(m)} \right)\right]
			\!\!{\pm}
			\cos\!\!\left[2\,\mathrm{Im}\!\!\left(\!\!\gamma^\ast\sum_{m=1}^N \alpha^{(m)} \right)\right]
		\right)\!\!.
	\end{aligned}
	\end{align}
\end{widetext}

	To apply our criteria in Eq. \eqref{eq:QQmulti}, and for simplicity, we set $\alpha^{(m)}_j=\alpha_j$ for all mode numbers $m$ and points in phase space, $\alpha_j$.
	In Fig. \ref{fig:OddCoh}, we exemplify the certification of nonclassicality for the state $|\Psi^{(-)}_{\gamma,3}\rangle$ [Eq. \eqref{eq:MultimodeState}] as a function of $\alpha_1=\mathrm{Re}(\alpha_1)$ and for a fixed $\alpha_2=0$.
	We remark that, for other mode numbers $N$, the plot looks quite similar.
	Most pronounced are nonclassical features for $\gamma$ close to zero, relating to a W state in which a single photon is uniformly distributed over three modes.
	For large $\gamma$ values, relating to a GHZ state, the negativities decrease, but $\det(M)$ remains below zero.
	We reiterate that our relatively simple, second-order correlations of $Q$ functions render it possible to certify the nonclassical properties of multimode, non-Gaussian states.

\subsection{Generalized phase-space representations and nonlinear detection model}\label{subsec:ExNL}
    
	For demonstrating how our phase-space matrix approach functions beyond $s$-parametrized distributions, we consider an on-off detector that is based on two-photon absorption \cite{JA69}.
	In this case, the POVM element for no click is approximated by
	\begin{align}
		\label{eq:NLdetector}
		\hat\Pi
		={:}e^{-\eta \hat n+\chi\hat n^2}{:}
		=\sum_{n=0}^\infty \frac{(2n)!}{n!}\left(\frac{\chi}{\eta^2}\right)^n
		{:}\frac{(\eta\hat n)^{2n}}{(2n)!}e^{-\eta\hat n}{:},
	\end{align}
	where ${:}(\eta\hat n)^{2n}e^{-\eta\hat n}/(2n)!{:}$ describes a measurement operator for $2n$-photon states with a linear quantum efficiency $\eta$.
	In this context, it is worth mentioning that $\chi\ll [e\eta^2]/[4n]$ has to be satisfied to ensure that the approximated POVM element correctly applies for photon numbers up to $2n$ \cite{comment:POVM}.
	The parameter $\chi$ relates to the nonlinear absorption efficiency.

	Based on such a nonlinear detector, we then define the non-Gaussian operator $\hat \Omega(\alpha;\eta,\chi)={:}e^{-\eta \hat n(\alpha)+\chi\hat n(\alpha)^2}{:}$, as described in Secs. \ref{subsec:Detect} and \ref{subsec:Regularized}.
	For a correlation measurement with two detectors (see Fig. \ref{fig:setup}), this then results in the correlation matrix elements $\langle{:}\hat\Omega(\alpha_i;\eta_i,\chi_i)\hat\Omega(\alpha_j;\eta_j,\chi_j){:}\rangle$.
	For specific parameters and up to a scaling with $\pi$, this correlation function also results in the nonlinear $Q_{\Omega}(\alpha)=\langle{:}\hat\Omega(\alpha;1,\chi)\hat\Omega(0;0,0){:}\rangle$ function (cf. Sec. \ref{subsec:Regularized} for the similarly defined $P_\Omega$), where $\sigma=\eta=1$.
	By extension, and using $\chi=\chi'$ and $\eta=1=\eta'$, these phase-space correlation functions also provide the entries required for the nonclassicality criterion.
	Here, it reads
	\begin{align}
		\label{eq:NLNC}
	\begin{aligned}
		\langle{:}\hat\Omega(\alpha_1;1,\chi)\hat\Omega(\alpha_1;1,\chi){:}\rangle
		\langle{:}\hat\Omega(\alpha_2;1,\chi)\hat\Omega(\alpha_2;1,\chi){:}\rangle
		\\
		-\langle{:}\hat\Omega(\alpha_1;1,\chi)\hat\Omega(\alpha_2;1,\chi){:}\rangle^2
		<0,
	\end{aligned}
	\end{align}
	which applies to the nonlinear detection scenario under study.

\begin{figure}
\centering
    \includegraphics[width=0.78\columnwidth]{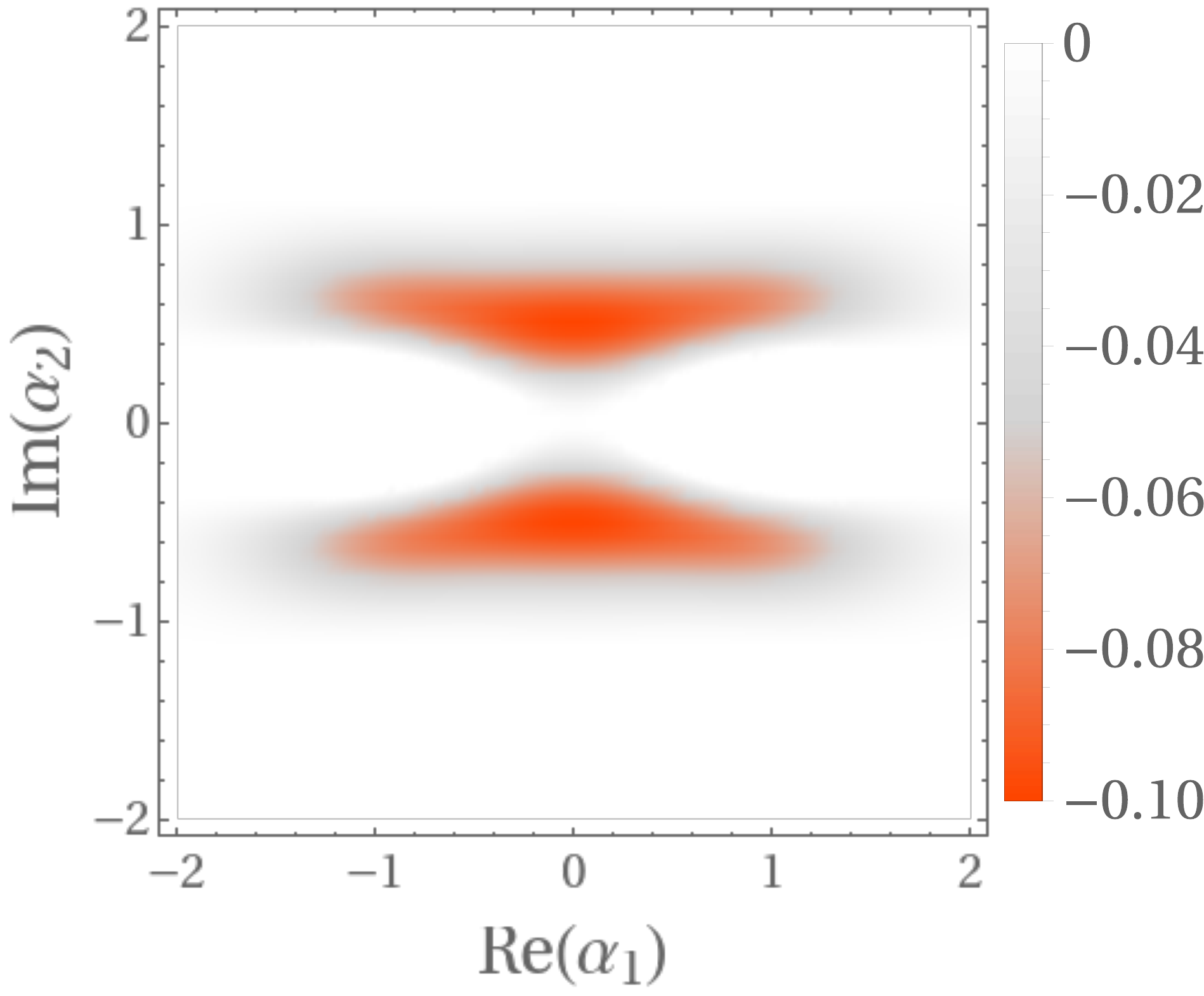}
	\caption{
		Application of the nonclassicality criterion in Eq. \eqref{eq:NLNC} as a function of $\mathrm{Re}(\alpha_1)$ and $\mathrm{Im}(\alpha_2)$, while fixing $\mathrm{Im}(\alpha_1)=\mathrm{Re}(\alpha_2)=0$.
		Nonlinear detectors---thus, a nonlinear $Q$ function---with $\eta=\sigma=1$ and $\chi=0.01$ are used, Eq. \eqref{eq:NLdetector}.
		Because of the negativities for the considered single-mode, symmetric state [cf. Eq. \eqref{eq:MultimodeState} for $N=1$ and $\gamma=1$], this state is shown to be nonclassical.
	}\label{fig:EvenCoh}
\end{figure}

	In Fig. \ref{fig:EvenCoh}, we apply this approach and consider the single-mode even coherent state $|\Psi^{(+)}_{\gamma,1}\rangle$ [cf. Eq. \eqref{eq:MultimodeState} for $N=1$], which is a non-Gaussian state, because we focused on the odd coherent state in the previous example.
	It is worth emphasizing that other methods to infer nonclassical light (e.g., the Chebyshev approach from Ref. \cite{BA19}) are incapable to detect this state's quantum features.
	Here, we can directly certify nonclassicality of this non-Gaussian state despite the challenge of also having a non-Gaussian detection model.

\section{Conclusion}\label{sec:Conclusion}

	In summary, we devised a generally applicable method that unifies nonclassicality criteria from correlation functions with quasiprobability distributions.
	Thereby, we created an advanced toolbox of nonclassicality tests which exploit the capabilities of both phase-space distributions and matrices of moments to probe for nonclassical effects.
	Furthermore, our framework is applicable to an arbitrary number of modes, arbitrary orders of correlation, and even phase-space functions perturbed through convolutions with non-Gaussian kernels.
	A measurement scheme was proposed to directly determine the elements of the phase-space matrix, the underlying key quantities of our method.
	In addition, we showed and discussed in detail that our treatment includes previous findings as special cases, is experimentally accessible even if other methods are not, and overcomes challenges of previous techniques when identifying nonclassicality.
    
    The phase-space-matrix approach incorporates nonclassicality tests based on negativities of the phase-space distributions, including the Glauber-Sudarshan $P$ function, and the matrix-of-moments approach as special cases.
    Thus, we were able to unify two major techniques for certification of nonclassicality.
    As the $P$ function and the matrix of moments themselves are already necessary and sufficient conditions for the detection of nonclassicality, the introduced phase-space-matrix approach obeys the same universal feature.
    In other words, for any nonclassical state there exists a phase-space matrix condition which certifies its nonclassicality.

	By applying our nonclassicality criteria to a diverse set of examples, we further demonstrated the power and versatility of our method.
	These examples covered discrete- and continuous-variable, single- and multimode, Gaussian and non-Gaussian, as well as pure and mixed quantum states of light.
	Remarkably, we used for all these states only the family of second-order correlations and phase-space distributions which are always nonnegative.
	Nevertheless, these basic criteria were already sufficient to certify distinct nonclassical effects on one common ground, further demonstrating the strength of our method.
	When compared to matrices of moments, the kinds of nonclassicality under study would require very different moments for determining the states' distinct quantum properties.
	Finally, we put forward an experimental scheme, only relying on readily available optical components, to directly measure the quantities required to apply our method.
	This scheme applies even for imperfect detectors with a nonlinear response.
	Furthermore, we want to add that the practicality and strength of the matrix of phase-space distributions in certifying nonclassicality of lossy and noisy quantum state can be experimentally demonstrated \cite{BBABZ20}.

	Here, we focused on nonclassical effects of light, owing to their relevance for photonic quantum computation and optical quantum communication.
	The introduced approach may be further developed for the certification of other quantum features, such as non-Gaussianity.
	Currently, our method detects nonclassicality for Gaussian and non-Gaussian states equally, which could be further developed for a more fine-grained quantumness analysis.
	However, instead of applying normal ordering, the construction of linear and nonlinear witnesses has to be adapted for this purpose.
	Furthermore, other kinds of quantum effects, such as entanglement, can be interpreted in terms of quasiprobabilities \cite{SW18} and are similarly witnessed through correlations \cite{HHHH09}.
	Thus, an extension to entanglement might be feasible as well.
	Therefore, our findings may provide the starting point for uncovering quantum characteristics through matrices of quasiprobabilities in other physical systems.
	Additionally, the derived framework can be utilized in the context of quantum information theory, such as in recently formulated resource theories of nonclassicality \cite{YBTNGK18,KCTVJ19} and other measures of nonclassicality \cite{TCJ20}, which employ the phase-space formalism \cite{SW18}, thus potentially benefiting from our phase-space correlation conditions for future applications.
	
	\paragraph*{Note added.}
	After finalizing this work, we have been made aware of a related work in preparation by J. Park, J. Lee, and H. Nha \cite{N20}.

\begin{acknowledgments}
	M.B. acknowledges financial support by the Leopoldina Fellowship Programme of the German National Academy of Science (LPDS 2019-01) and by the Erwin Schr\"odinger International Institute for Mathematics and Physics (ESI), Vienna (Austria) through the thematic programme on \textit{Quantum Simulation - from Theory to Application}.
	E.A. acknowledges funding from the European Union's Horizon 2020 research and innovation programme under the Marie Sk\l{}odowska-Curie IF InDiQE (EU project 845486).
	The authors thank J. Park, J. Lee, and H. Nha for valuable comments.
\end{acknowledgments}



\begin{thebibliography}{99}
\bibitem{comment:3x3}
		The determinant of a $3\times 3$ matrix $X=\left(\begin{smallmatrix}
			\mu & u & v
			\\ u & U & \chi
			\\ v & \chi & V
		\end{smallmatrix}\right)$ takes the general form $\det X=[(\mu U-u^2)(\mu V-v^2)-(\mu\chi-uv)^2]/\mu$, which is particularly interesting for the case $\mu=1$ because it relates to cross-correlation functions.
\bibitem{comment:derivatives}
        It is worth noting that, in quantum optics, the partial derivative with respect to a complex amplitude $\alpha$ is given in terms of partial derivatives of the real and imaginary part, $\partial_\alpha=(\partial_{\mathrm{Re}(\alpha)}+i\partial_{\mathrm{Im}(\alpha)})/2$ and $\partial_{\alpha^\ast}=(\partial_{\mathrm{Re}(\alpha)}-i\partial_{\mathrm{Im}(\alpha)})/2$.
\bibitem{comment:darkcount}
	    For simplicity, we assume an equal dark-count rate $\delta$ for both detectors.
	    However, one can readily generalized this to different dark-count rates for each detector, as $\det (M)<0$ remains a sufficient nonclassicality condition.
\bibitem{comment:POVM}
	    The approximate POVM element in Eq. \eqref{eq:NLdetector} has a decomposition in terms of lossy even photon-number operators with the expansion coefficients $[(2n)!/n!](\chi/\eta^2)^n$, which diverge for $n\to\infty$.
        Using the bounds $\sqrt{2\pi}m^{m+1/2}e^{-m}\leq m!\leq e m^{m+1/2}e^{-m}$, one finds the bound $\chi\ll e\eta^2/[4n]$ to satisfy $[(2n)!/n!](\chi/\eta^2)^n\leq [e/\sqrt\pi]([4n\chi]/[e\eta^2])^n\leq 1$ for correctly applying this approximation for upto $2n$ photons.
        Also note that for coherent states, one obtains the nonnegative function $\langle\alpha|\hat\Pi|\alpha\rangle=\exp(-\eta|\alpha|^2+\chi|\alpha|^4)\geq0$, representing the non-Gaussian integration kernel $\Omega$.
\end{thebibliography}
\end{document}